# Magnetically Tunable Chirality in 2D Liquid Crystalline WS$_2$ Nanocomposites

*Benjamin T. Hogan[1], Yulia Gromova[2], Evgeniya Kovalska[1], Alexander Baranov[2], Monica F. Craciun[1] and Anna Baldycheva[1*].*

1 University of Exeter, North Park Road, Exeter, UK, EX4 4QF.

2 ITMO University, Kronverksky Pr., St. Petersburg, 197101, Russia

The first observation of tungsten disulfide liquid crystalline nanocomposites in dispersions of liquid phase-exfoliated flakes is demonstrated in a range of organic solvents. The nanocomposites demonstrate significant birefringence and reconfigurable optical chirality as observed in the linear and circular dichroism measurements respectively. Under an applied magnetic field of ±1.5T the chirality can be switched ON/OFF, while the wavelength range for switching can be tuned from large to narrow range by the proper selection of the host solvent. In combination with photoluminescence capabilities of WS$_2$, this opens a pathway to a wide variety of applications, such as deposition of highly uniform films over large areas for photovoltaic devices, as shown here.

Two-dimensional (2D) nanocomposite materials with dynamically tunable properties are currently emerging as a highly-promising- and highly-desirable- class of novel functional materials, owing to the diverse and extraordinary properties that they can possess[1–3]. By producing liquid crystalline nanocomposites based on 2D materials, structural reconfigurability with associated tailoring of properties has recently been achieved[4–9]. However, until now, such liquid crystal phases[2] have been reported predominantly for graphene oxide dispersions[10–16] with some limited reports for carbon allotropes such as graphene[7,17,18] and carbon nanotubes[6] as well as for one transition metal dichalcogenide (TMD), molybdenum disulfide[8]. Comparing these examples to the breadth of two-dimensional materials known, one can immediately see the limitations of the range of materials explored. Of significant intrigue are reports that such liquid crystalline phases can also show chirality[17,19] as reconfigurable chiral materials are in strong demand for photonics applications[20,21]. One of the most desirable methods for chiral material reconfiguration is magnetic tuning, due to the possibility for either *in-* or *ex-situ* switching with favourable power requirements[22,23].

Of particular interest, tungsten disulfide ($WS_2$) is a TMD with significant potential for applications in a variety of areas. Its various exciting properties have led to widescale integration of $WS_2$ in heterostructures and junctions[24–28] amongst other applications such as field effect transistors[27], hydrogen evolution[29], saturable absorbers[30], supercapacitors[28] and battery anodes[31]. In particular, strong room temperature photoluminescence makes it an ideal candidate for applications in solar power[24,32], for instance, while second harmonic generation allows use in non-linear photonic devices[26].

In its bulk crystalline form, $WS_2$ consists of dark grey hexagonal crystals built up of layered sheets held together by van der Waals forces. Within each individual sheet (Fig. 1a), the tungsten has a trigonal prismatic co-ordination geometry, while the sulfur has a pyramidal co-ordination, such that each sulfur bridges between three tungsten centres, whilst each tungsten is bonded to six sulfurs. Dating back 60 years, the bulk crystalline form found applications as a catalyst for numerous reactions[33,34]. More recently, significant interest was generated from the discovery that $WS_2$ can form nanotube structures analogous to carbon nanotubes, representing the first inorganic nanotubes to be discovered[35]. However, with the current explosion in the investigation of 2D materials, it is the van der Waals layered nature of $WS_2$ that generates the greatest interest[3,36]. The weak nature of the interlayer bonding makes it a prime candidate for exfoliation to few- or single-layered flakes of high aspect ratio. Whereas bulk $WS_2$ is an indirect bandgap semiconductor, with a bandgap of around 1 $eV$, monolayers possess a direct bandgap of 1.8 $eV$. This direct bandgap presents a significant advantage for applications over graphene which has no bandgap.

Herein, we demonstrate- for the first time- the possibility of a liquid crystalline phase of dispersions of $WS_2$ particles (Fig. 1b) in a variety of solvents. Of particular interest is the emergence under applied magnetic field, or without applied field, of circular dichroism of dispersions in particular solvents (Fig. 1c). This observation is particularly unexpected due to the achiral nature of both $WS_2$ and the solvents used. As such we demonstrate a reconfigurable chiral material that can be readily integrated or applied within photonic and optoelectronic devices by a variety of different means.

Starting from bulk $WS_2$ particles, with dimensions around a few microns, dispersions were produced in a range of solvents, with a range of concentrations. To accurately compare the effect of different solvents, it was necessary to have homogeneous particle size distributions between samples. Hence, an initial solution was prepared with IPA as the solvent. To break down the material a process of ultrasonication (see Supplementary Methods) was used with 5 hour-long periods separated by 30 minutes each to prevent excessive heating of the solvents. The resultant dispersions were then put through a process of centrifugation to remove residual bulk material and narrow the distribution of particle sizes present in the solutions and fractioned to ensure only suitably exfoliated particles remained. The resultant solution was then dried under vacuum to fully remove the solvent, before being re-dispersed in the required solvents for the final solutions. After re-dispersion, the solutions were again ultrasonicated (for a few minutes) to prevent any aggregated exfoliated particles remaining in the solutions. As the concentration is changed significantly following the centrifugation, it is necessary to re-establish the concentration following that step. The re-dispersion process also allowed for accurate knowledge of the concentrations of the solutions.

Schlieren-like textures can be observed in the liquid composites without the need for polarising optics with bright and dark regions corresponding to changes in reflectance of the sample due to alignment of the particles relative to the incident light used for imaging- shown here for dispersions in isopropanol- (Fig. 2) across a range of concentrations, from of $5mg.mL^{-1}$ to $0.1mg.mL^{-1}$, with textures observed for all concentrations except the lowest. Similar textures are observed in other solvents. Under cross-polarised optical microscopy, after suitable control of the concentration by solvent evaporation, liquid crystallinity was observed (Supplementary Figure 1) as demonstrated by optical birefringence.

In particular, samples in tetrahydrofuran (THF), isopropanol (IPA) and chloroform demonstrated promising properties. The linear dichroism of dispersions in THF, IPA and chloroform was measured in the visible range (200-800 *nm*). For these measurements, a quartz cuvette with a path length of 2mm was filled with the dispersion and the transmission of light through it analysed. Bulk $WS_2$ has no dichroic response, however dichrism can be induced after exfoliation due to the shape anisotropy of the resultant board-like particles (see Supplementary Information). In an isotropic distribution of particles in solution, no dichroism would be observed as the orientation of the particles would be random. However, where there is an ordering of the particles relative to one another, a dichroic response can be observed; such ordering is the principle property determining liquid crystallinity. The dichroism is a product of the different effective refractive indices produced by particles being aligned with their short axes either parallel or perpendicular to the polarisation of the incident light. The dichroism is induced by self-assembly in the solvent, therefore we look at the composite effects of both the solvent and the dispersed particles. As such, the absorbance of the solvents below 225nm (See Supplementary Results) may mask any dichroism. However, from a material application viewpoint, the dichroism can therefore be considered negligible in this region in any case.

As seen in Figure 3a, linear dichroism can be observed across a broad range of wavelengths for dispersions in all three solvents. For these measurements, a concentration of $1mg.mL^{-1}$ was used for THF and a concentration of $5mg.mL^{-1}$ for IPA and chloroform. However, there is expected to be a significant concentration dependence of the liquid crystallinity. Hence, solutions with a range of concentrations were measured, to establish where the onset occurs. Significant increases can be observed in the linear dichroism of solutions in IPA (Fig. 3b) and chloroform (Supplementary Fig. 2) as the concentration is increased from $0.1\ mg.mL^{-1}$ to $5\ mg.mL^{-1}$. However, no clear transition concentration from isotropic to liquid crystalline is observed. This is likely due to the non-homogeneous distribution of particle sizes present in the solution, leading to a broad region of biphasic stability where liquid crystallinity occurs in localised isolation rather than throughout the entirety of the solution. The effect of the particle size was investigated by comparing solutions of exfoliated particles, with those of unexfoliated bulk material which still possessed some shape anisotropy. Unexfoliated particles had typical sizes of 1-10µm, with the mean size around 2.5x2.5µm, determined from optical microscopy. Thickness was determined to be predominantly bulk-like from Raman spectroscopy, although determination of any shape anisotropy was not possible with Raman spectroscopy for which accurate determination of thickness is limited beyond ~10 layers. Exfoliated particles had typical dimensions of around 500nm up to a few microns in length and width (as observed in microscopy characterisation) with mean size around 1x1µm, and typical thicknesses of 1-10 layers as observed in Raman measurements. However, characterisation also revealed the presence of occasional much larger particles with bulk like characteristics, and some particles <500nm in length or width. It was also observed that there were significant differences in the shapes of the exfoliated particles produced. For lower concentrations, little difference is observed between the exfoliated and unexfoliated samples. However, as the concentration was increased, significant increases in the linear dichroism were observed for the exfoliated and unexfoliated solutions, such that the exfoliated solution showed much stronger dichroism than the unexfoliated (Fig 3c). For all samples, nanoparticle precipitation partially occurred due to the containment in the confined geometry of the cuvette, as observed from sedimentation after measurements. The largest flakes are likely to precipitate first, and these are expected to have the least effect on the liquid crystallinity. With precipitation, a trend towards the pure solvent spectrum would be expected and hence for the spectra to tend towards zero after normalisation. This is not observed in the spectra recorded, hence there is negligible effect of the precipitation on the dichroism results obtained.

The circular dichroism of the samples was also analysed for the solutions in THF, IPA and chloroform. As can be seen in Figure 4, circular dichroism was observed for all three solvents. Similar circular dichroism has previously been observed for liquid crystals of graphene oxide[19], where a twisted lamellar structure is proposed (Fig. 1c). In this structure, sheets of the 2D material form lamellar blocks of concentration dependent sizes. Separate blocks then arrange with a helical pitch between them, giving rise to a twisted ordering over long ranges; thus leading to circular dichroism. However, this mechanism is currently both disputed and unproven[2]. Here, the inherent noise in the measurements due to absorption and scattering makes any qualitative analysis difficult. Whereas molecular chirality is related directly to the electronic transitions as seen in the absorption spectra of the molecule, for supramolecular chiral assemblies the chirality is related to the helical pitch of the structures- hence the chirality is unrelated to the absorption band of tungsten disulfide (measured in solution at 635nm). By applying a magnetic field to the solutions during measurement, significant enhancement of the circular dichroism is observed for THF and IPA, with a clear broad band emerging for wavelengths >500nm. Under applied magnetic field, especially for IPA, the circular dichroism can be enhanced due to an aligning force giving a preferred directional axis for the helicity. Broad widths of the circular dichroic bands are expected due to the non-homogeneity of the particle sizes present in the solution.

Additionally, for chloroform, under the applied field a very strong peak in the circular dichroism is observed at around 230nm. This peak switches sign with the applied field. The peak can also be observed in the pure solvent, however, as results are normalized versus the pure solvent spectrum there is enhancement observed due to the presence of the dispersed nanoparticles. However, it is the properties of the composite liquid crystalline system that are of interest, hence it is justifiable that the composite system of tungsten disulfide dispersed in chloroform represents a magnetically reconfigurable chiral material.

While different solvents can affect the rate, efficiency and quality of particles obtained by liquid phase exfoliation, due to the fact that controlling the concentration generally relies on drying and redispersion the solvent choice for obtaining a liquid crystalline state can be independent of that for exfoliation. The different properties achievable with different solvents show that there is no single definable 'best' solvent for use. Rather, the choice of solvent should be application driven. For example, to achieve reconfigurable chirality at 750nm, IPA would be the best choice, whereas at 230nm chloroform is preferable. The single condition is that $WS_2$ should be soluble enough to allow dispersion at the necessary concentrations, which is broadly true for all organic solvents.

To demonstrate one area where the liquid crystalline solutions could find application, solutions were filtered to produce homogeneous thin films and stacked layers of few-layer $WS_2$. These thin films were then readily transferred to substrates (Fig. 5a-c), marking a facile and scalable method towards large scale integration of the two-dimensional material. Films transferred from the LC state show far greater homogeneity than those from non-LC solutions (Fig. 5c). By using liquid crystalline solutions for filtering, as opposed to non-liquid crystalline solutions, a significant increase in the homogeneity of the deposited films is observed. Importantly, individual flakes of $WS_2$ exfoliated by liquid phase exfoliation can be photoluminescent when deposited from the liquid crystal state (Fig. 5d), as is also observed for flakes produced by other methods. After transfer to a substrate, Raman spectroscopy mapping (Fig. 5e-f, Supplementary Fig. 3) of the surface shows complete coverage across large areas, with the variation in signal owing to differing numbers of layers present. The ability to cover large areas with few- or monolayer $WS_2$ opens up great possibilities in photovoltaic devices, owing to the photoluminescence of $WS_2$ in the visible range.

This work demonstrates that using only standard liquid phase exfoliation processes (ultrasonication, centrifugation and control of concentration) a liquid crystalline state of $WS_2$ particles can be achieved in a range of solvents. The discovery that $WS_2$ LC dispersions can simultaneously demonstrate reconfigurability under applied magnetic field and chirality, while also combining the intrinsic optical and electrical properties of the 2D $WS_2$ flakes is of significant interest, opening the way towards countless applications in opto-electronics and photonics. Future optimisation and tuning of the synthesis holds great potential for the generation of a new class of true reconfigurable '*wonder materials*'. The discovery of this novel reconfigurable phase could lead to a marked expansion in the uptake of $WS_2$ for device based applications, due to the increased scope for scalable integration with reduced drawbacks such as time, cost and quality as well as the significant boost in the range of potential uses with the observation of reconfigurable chirality.

## ASSOCIATED CONTENT

**Supporting Information**. Supplementary discussion of methods and theory, supplementary results are contained in the following file, available free of charge: Supplementary information (PDF).

## AUTHOR INFORMATION

**Corresponding Author**

* a.baldycheva@exeter.ac.uk

**Author Contributions**

The manuscript was written through contributions of all authors. All authors have given approval to the final version of the manuscript.

ACKNOWLEDGMENT

We acknowledge financial support from: The Engineering and Physical Sciences Research Council (EPSRC) of the United Kingdom via the EPSRC Centre for Doctoral Training in Electromagnetic Metamaterials (Grant No. EP/L015331/1) and via Grant Nos. EP/N035569/1, EP/G036101/1 and EP/M002438/1, and the Royal Society International Exchange Grant 2016/R1. Additionally, Y.G. and Al.B. thank the Russian Foundation for Basic Research (Grant No. 17-52-50004) for partial financial support.

ABBREVIATIONS

2D, two-dimensional; TMD, transition metal dichalcogenide; $WS_2$, tungsten disulfide; IPA, isopropanol; THF, tetrahydrofuran; LC, liquid crystal.

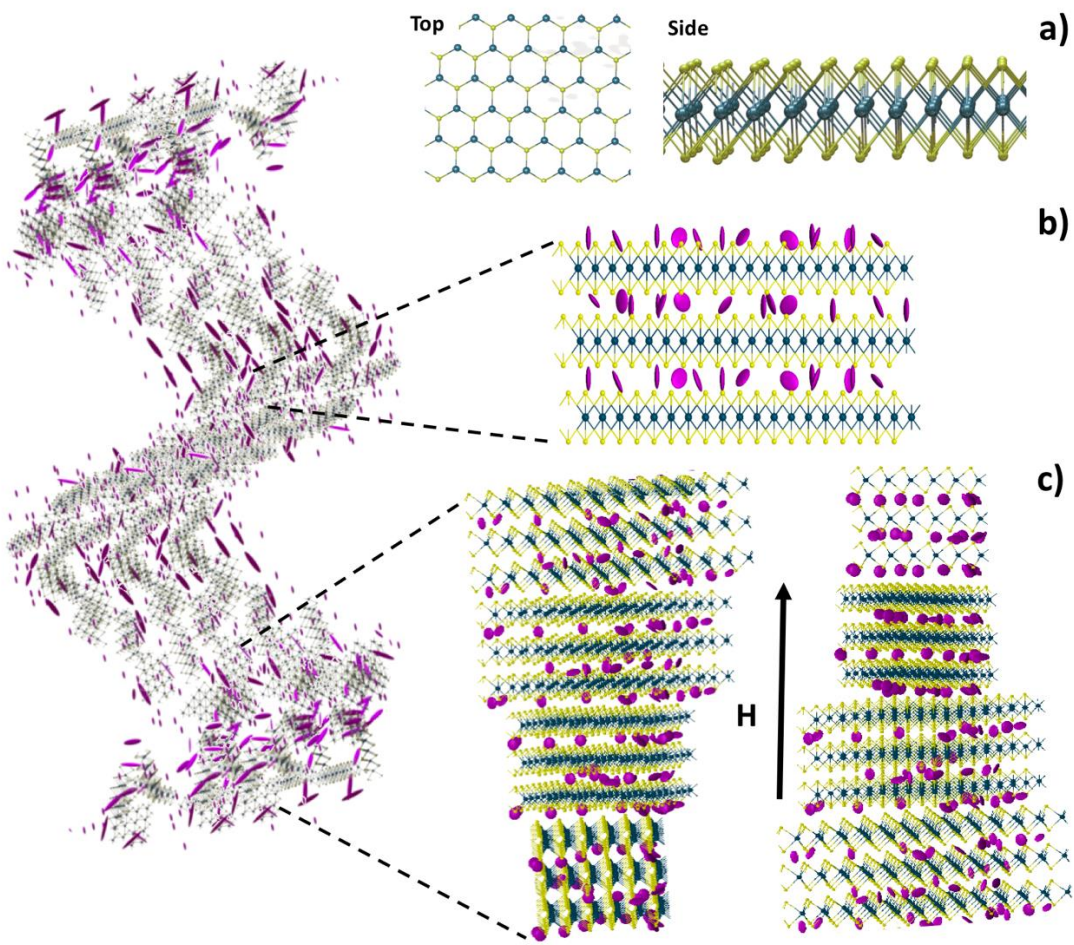

**Figure 1:** Schematics representations of liquid crystalline dispersions of tungsten disulfide. **a)** Top and side views of the bonding structure of tungsten disulfide. Tungsten atoms are represented in blue and sulfur in yellow. **b)** Liquid crystalline assembly of monolayers of tungsten disulfide, separated by intercalating solvent molecules (purple), retaining crystalline ordering between sheets. **c)** A structure of the chiral self-assembly of tungsten disulfide liquid crystalline dispersions under applied magnetic field.

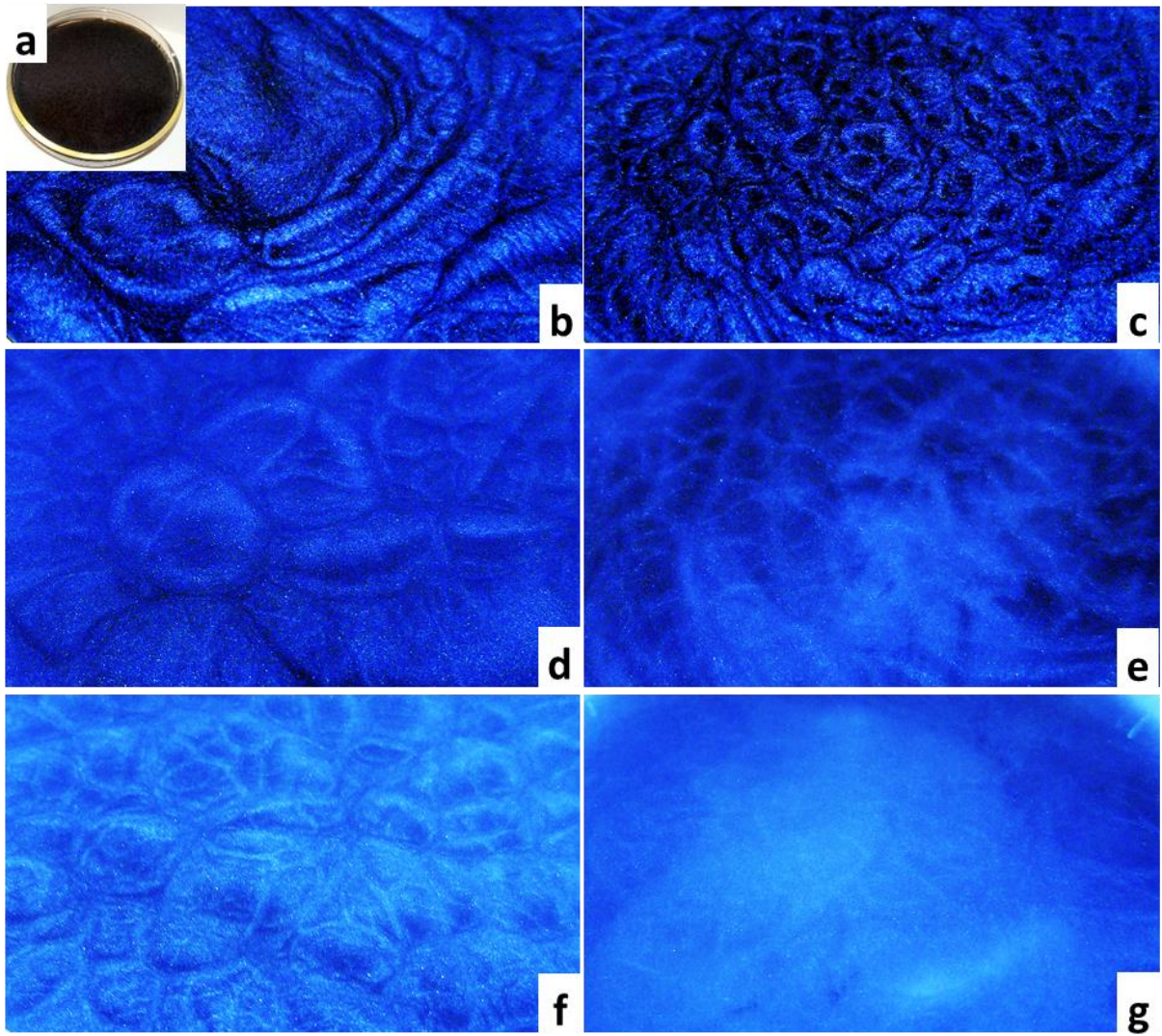

**Figure 2: a-g)** Optical images of $WS_2$ dispersions with clear bright and dark states visible depending on flake domain orientation relative to the incident light; **a)** shows the texture across a large surface of the dispersion, **b-g)** give a closer look at the surface for concentrations of 5mg.mL$^{-1}$, 2.5mg.mL$^{-1}$, 1mg.mL$^{-1}$, 0.5mg.mL$^{-1}$, 0.25mg.mL$^{-1}$, and 0.1mg.mL$^{-1}$ respectively.

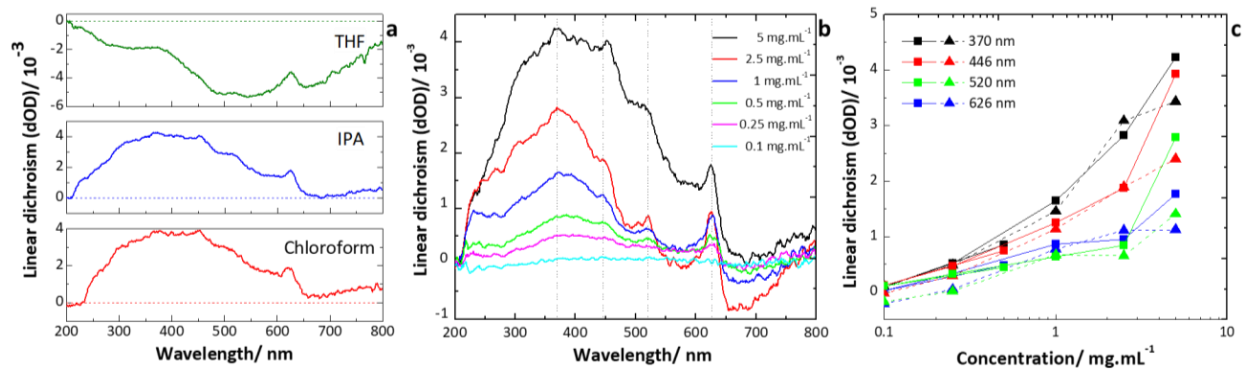

**Figure 3: a)** Linear dichroism of WS$_2$ dispersions in THF (1mg.mL$^{-1}$), IPA (5mg.mL$^{-1}$) and chloroform (5mg.mL$^{-1}$). **b)** Linear dichroism of dispersions of WS$_2$ in IPA at different concentrations: 5mg.mL$^{-1}$ (black); 2.5mg.mL$^{-1}$ (red); 1mg.mL$^{-1}$ (blue); 0.5mg.mL$^{-1}$ (green); 0.25mg.mL$^{-1}$ (magenta); 0.1mg.mL$^{-1}$ (cyan). **c)** Linear dichroism of dispersions of WS$_2$ in IPA at specific wavelengths: 370 nm (black); 446 nm (red); 520 nm (green); 626 nm (blue). Solid lines (and squares) are for samples that had been exfoliated as described whereas dashed lines (and triangles) are for dispersions of much larger (unexfoliated) particles. Error bars are smaller than the symbol size.

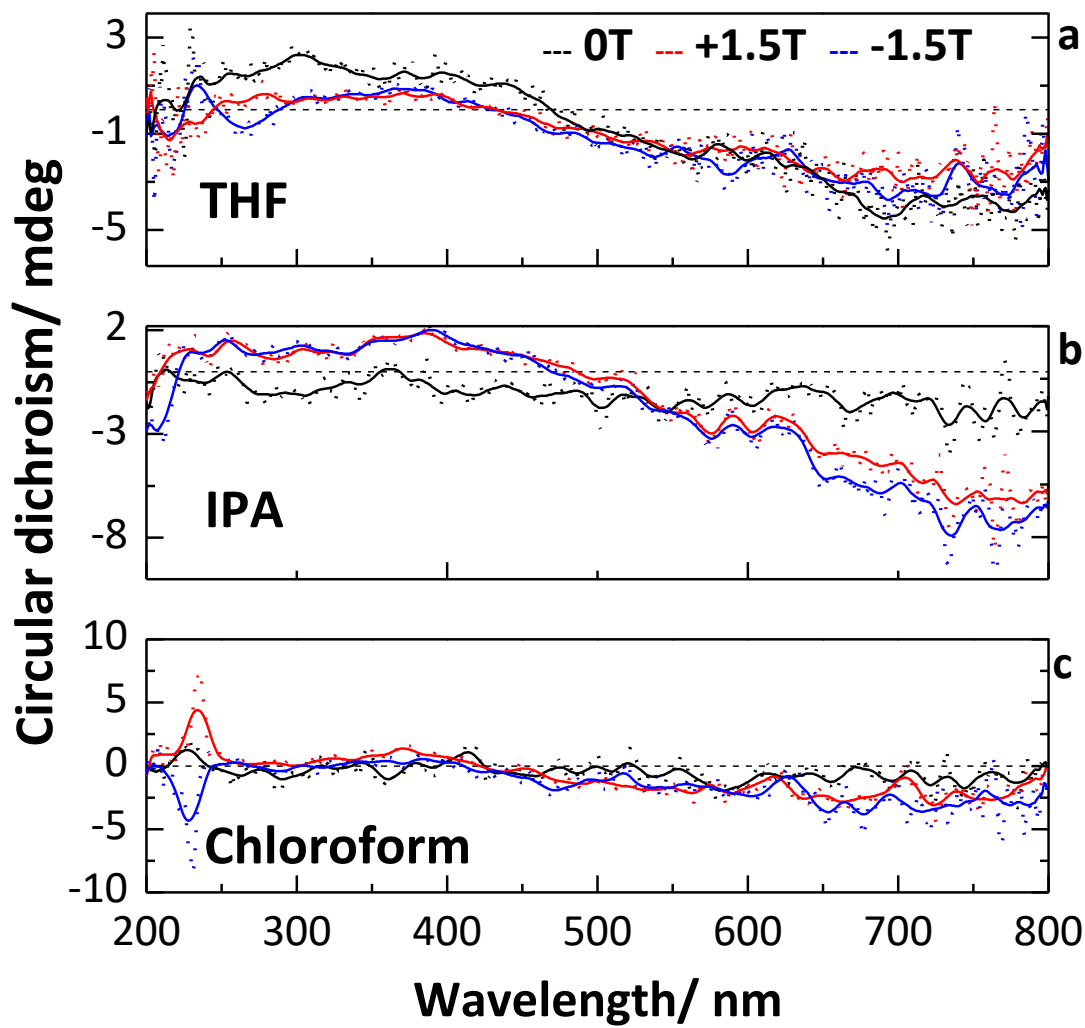

**Figure 4:** Circular dichroism of dispersions of WS$_2$ in different solvents: **a)** THF, **b)** IPA and **c)** chloroform, under applied fields of 0T (black), +1.5T (red) and -1.5T (blue). Dashed lines are normalised data including noise, solid lines give an indication of the general trend of the circular dichroism accounting for the noise.

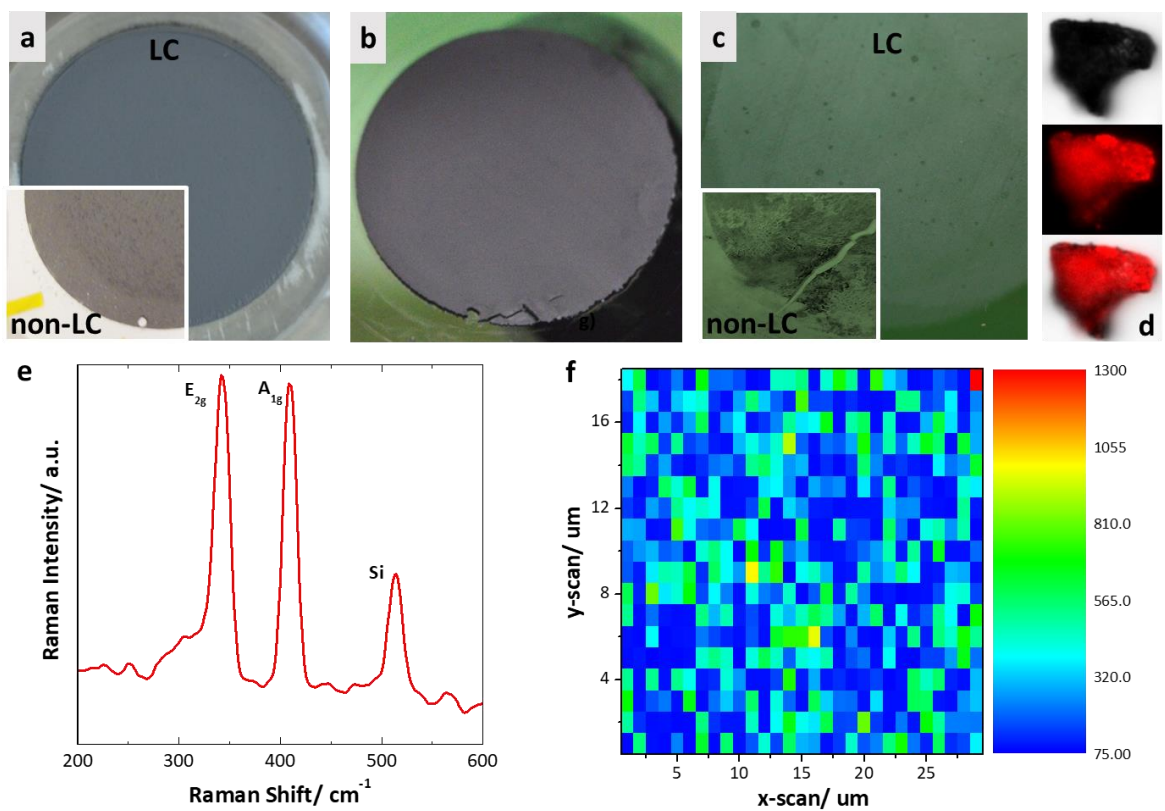

**Figure 5: a-c)** Progression of a dried film produced from a liquid crystalline dispersion of tungsten disulphide: **a)** deposition onto filter from an LC solution, inset- a deposition onto the filter from a non-LC solution, **b)** the resultant film; **c)** a considerably thinner film produced by the same method from the liquid crystalline phase, inset- the film produced from the non-LC filtrate. **d)** Photoluminescence of a single representative flake from the liquid crystalline phase. **e)** Raman spectrum of few-layer $WS_2$ showing the two expected peaks in addition to a silicon peak. **f)** Raman map of the $E_{2g}$ peak of $WS_2$ showing coverage across the whole surface for an area of 29 μm x 18 μm.